\documentclass[aps, pre, twocolumn]{revtex4-1}
\usepackage{graphicx, amsfonts, amssymb, amsmath}
\usepackage{bm}
\usepackage{times}
\usepackage[colorlinks,citecolor=blue,linkcolor=red]{hyperref}
\usepackage{color}

\begin{document}
\title{Anomalous Transient Heat Conduction in Fractal Metamaterials}

\author{Wuxi Lin}
\affiliation{%
Center for Phononics and Thermal Energy Science, China-EU Joint Lab on Nanophononics, Shanghai Key Laboratory of Special Artificial Microstructure Materials and Technology,
School of Physics Sciences and Engineering, Tongji University, Shanghai 200092, China
}%
\author{Shengpeng Huang}
\affiliation{%
Center for Phononics and Thermal Energy Science, China-EU Joint Lab on Nanophononics, Shanghai Key Laboratory of Special Artificial Microstructure Materials and Technology,
School of Physics Sciences and Engineering, Tongji University, Shanghai 200092, China
}%
\author{Jie Ren}
\email{Corresponding address: Xonics@tongji.edu.cn}
\affiliation{%
Center for Phononics and Thermal Energy Science, China-EU Joint Lab on Nanophononics, Shanghai Key Laboratory of Special Artificial Microstructure Materials and Technology,
School of Physics Sciences and Engineering, Tongji University, Shanghai 200092, China
}%


\begin{abstract}
Transient dynamics of heat conduction in isotropic fractal metamaterials is investigated. By using the Laplacian operator in non-integer dimension, we analytically and numerically study the effect of fractal dimensionality on the evolution of the temperature profile, heat flux and excess energy under certain initial and boundary conditions.  
Particularly, with randomly distributed absorbing heat sinks in the fractal metamaterials, we obtain an anomalous non-exponential decay behavior of the heat pulse diffusion. and an optimal dimension for efficient heat absorption as a function of sink concentrations.
Our results may have potential applications in controlling transient heat conduction in fractal media, which will be ubiquitous as porous, composite, networked, hierarchical meta-materials.
\end{abstract}

\maketitle

\section{Introduction}
Heat problem has attracted lots of attention in meta-structures in the past decade~\cite{jphuang, ny1, ny2, ny3}.
Particularly, heat propagation as well as mass or excitation diffusion in fractal media are of great importance in our everyday life~\cite{cite0}. This is because fractal meta-media describe the porous, composite, networked, hierarchical metamaterials, in which a part of the structure resembles larger entities or the whole structure. Such self-affine structural patterns
are ubiquitous in many fields such as physics, material science and life science. For example, in branching artery network~\cite{cite3}, in photosynthesis~\cite{cite2}, and in bones~\cite{bone}, the fractal media of statistical self-similarity is quite involved. Therefore, fractal metamaterials have attracted much attention across diverse research fields, ranging from mechanics~\cite{mechanics1,mechanics2} and elastics~\cite{elastic}, to acoustics~\cite{acoustic1,acoustic2,acoustic3} and optics~\cite{optics1,optics2,optics3}. Different from those research that focused on effects of fractal structure on wave dynamics, heat conduction and transfer in fractal metamaterials is in general a diffusion process.

Effective thermal conductivity is often used to characterize the heat diffusion in fractal media of fractal dimensions. Various methods have been developed to investigate the effective thermal conductivity of porous media~\cite{eff1, eff2,eff3,eff4,eff5,eff6,eff7,eff8}, that may be even used to build a thermal diode~\cite{holydiode}. Pitchumani~\cite{cite4} applied fractal theory in the research of the effective thermal conductivity of unidirectional fibrous composites. Yu and Cheng~\cite{cite5} developed a fractal model to calculate the effective thermal conductivity of mono- and bi-dispersed porous media. Using thermal-electrical analogy, Yu~\cite{cite6} and Kou~\cite{cite7} presented fractal models and fractal analysis of effective thermal conductivity of composites with embedded fractal-like tree networks and saturated fractal porous media, respectively.

To study fractal media of fractal dimensions, two methods are often applied: one is fractional calculus~\cite{fracint1,fracint2} and the other one is calculus in fractional dimension space. In fractional calculus, factorial is replaced by gamma function to expand the application scope of previous calculuses~\cite{cite8}, where the differential and integral calculus of time are usually involved. While calculus in fractional dimension space pays more attention to the geometric property of the non-integer space. An axiomatic system is established first~\cite{cite1} and then applied for excitons in fractional dimensional space~\cite{He}, and sequentially generalized~\cite{JPA, cite9}. Stillinger constructed the axiomatic basis for spaces with non-integer dimension, and gave the form of Laplacian operator in fractal space~\cite{cite1}. Tarasov suggested a generalization of vector calculus for the case of non-integer dimensional space~\cite{cite9}, and gave a solution of heat propagation in fractal pipe and rod under cylindrical coordinate~\cite{cite10}. 

Relevant experiments have also been carried to study heat conduction in fractal media. Rozanova-Pierrat investigated how the shape of a prefractal radiator may enhance global heat transfer at short time~\cite{cite11}. Cervantes-Alvarez reported the thermal characterization of plate-like composite samples made of polyester resin and magnetite inclusions~\cite{cite12}.


\begin{figure}
\vspace{2mm}
\scalebox{0.36}[0.34]{\includegraphics{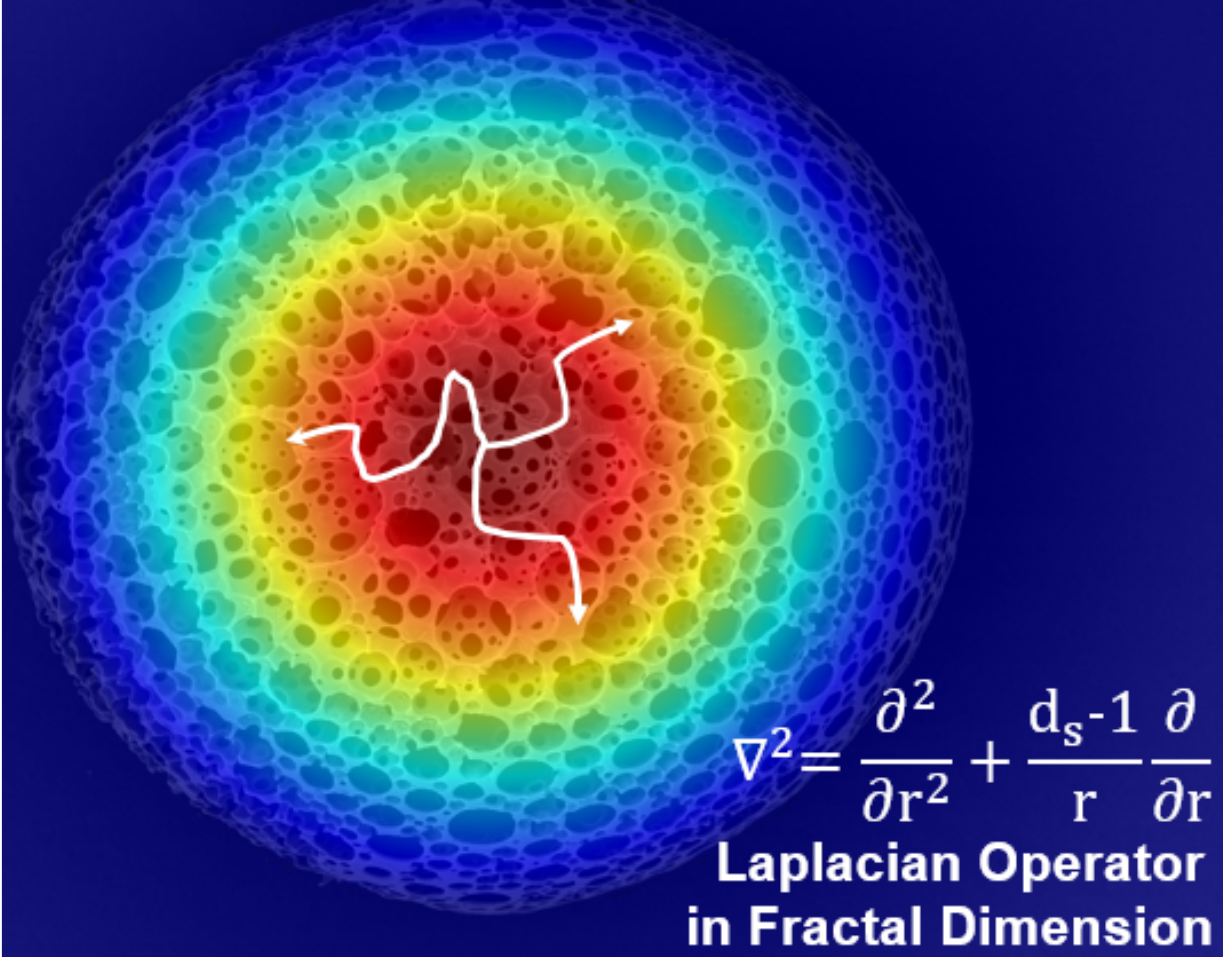}}
\vspace{-2mm}
\caption{Schematic illustration of the transient heat conduction in fractal meta-media, which can be porous, composite, networked materials, showing a self-affine pattern that a part of the structure resembles larger entities or the whole structure. The heat diffusivity is described by the Laplacian operator in fractional dimension.} 
\label{fig:Fig1}
\end{figure}


In this paper, we use calculus in fractional dimension to describe transient heat conduction in isotropic fractal media under specified boundary and initial conditions. For simplicity, we consider spherically symmetric media space. The initial temperature distribution in the spherical media is arbitrary and the boundary temperature keeps constant. We thus analytically and numerically study the evolution of the temperature profile, heat flux and excess energy. 
In particular, we analytically obtain the non-exponential decay of the heat pulse diffusion in the fractal media with randomly distributed absorbing heat sinks. 
In this case an optimal dimension exists for faster heat absorption, which depends on the heat sink concentrations.

However, we have to mention that the mathematical operators in fractional dimension space may not fully describe the real fractal media, because the mathematics here does not maintain the strict self-similarity at all scales. Nevertheless, it captures the main characteristic of fractal media in large scale in the sense of effective media. For this reason, calculus in fractional space can offer us important insights about the diffusion behavior in the fractal media. Moreover, when varying dimensions, we consider the case of constant mass density following the same spirit as previous~\cite{cite9, cite12}. Our results and formulas can be readily extended to the case of varying mass density due to different arrangements of composite units at different dimensions.

\section{Theory of the fractal heat conduction}
In this section, we will provide the basic theory of the heat conduction in a fractal material. 
Although Fourier's law may break down at nanoscale~\cite{cite21}, throughout this work the fractal structures of interest are beyond the mesoscopic level and the unit size of the fractal structure is above the micrometer, so that Fourier's law is valid. 
We consider the heat conduction in homogenous isotropic media, which is described by the Fourier's law:
$\vec{J}=-\kappa\nabla{T}$,  
with $\vec{J}$ the heat flux vector, $\kappa$ the thermal conductivity of the medium, $T$ the temperature as a function of location. The continuous equation of energy conservation requires the in-out flux balance:
\begin{equation}
\oint_{V} f(\vec{r},t)dV -\oint_{S} \vec{J}(\vec{r},t)\cdot d\vec{S}\ = \frac{\partial }{\partial t}\oint_{V} c \rho T(\vec{r},t) dV
\label{eq:conversation of energy}
\end{equation}
where  $f(\vec{r},t)$ denotes the intensity of the internal heat source, $c$ is the heat capacity, $\rho$ is the medium's mass density, $S$ and $V$ denotes the surface and volume of the medium, respectively. As such, $\nabla \cdot\vec{q}+\rho c \frac{\partial T}{\partial t}=f(r,t)$. Considering both the divergence theorem and Fourier's law, we arrive at the differential form: 
\begin{equation}
\nabla^2{T}(\vec{r},t)-\frac{1}{D}\frac{\partial T(\vec{r},t)}{\partial t}=-\psi(\vec{r},t),
\label{common equation}
\end{equation}
where $\psi(\vec{r},t)={f(\vec{r},t)}/{\kappa}$ is the source term and $D={\kappa/(c\rho)}$ has the physical meaning of heat diffusion coefficient. 
The Laplacian operator in the fractal dimension can be written as
\begin{equation}
\nabla^2=\frac{\partial^2}{\partial r^2} +\frac{d_s-1}{r}\frac{\partial}{\partial r} +\frac{\frac{\partial}{\partial \phi}(\sin\phi^{n-2}\frac{\partial}{\partial \phi})}{r^2\sin\phi^{n-2}}+\frac{\frac{\partial^2}{\partial \phi^2}}{r^2\sin\phi^2},
\label{Laplacen operator in D-dimension media}
\end{equation}
where the non-integer $d_s\geq 1$ denotes the fractional dimension. This is a very important result cited from F. H. Stilinger's original work~\cite{cite1} [see also, the Laplace-Beltrami operator]. Stilinger's work is a very important achievement in the area of fractal geometry. He establishes an axiomatic system under the condition that the space's dimension is a fraction and on the basement he derives expressions of many basic geometry quantities such as the length and the volume element. On the basement  he provides the expression of the Laplacian operator in the fractional dimension space. His results are widely cited in the papers about the researches of the physical characteristics in fractal media. For example, 
Tarasov has referred to his theory to solve the problem about heat transfer in fractal media~\cite{cite10}, and Wei-Ping Zhong et al. have used his theory to solve the problem about spatiotemporal accessible solitons in fractional dimensions~\cite{cite14}. 

In view of the fact that the homogenous isotropic thermal medium is spherically symmetric, all physical quantities can be represented as scalar functions of the radius distance, such as $T(\vec{r},t)=T(r,t), \psi(\vec{r},t)=\psi(r,t)$.
In such fractal media, the Laplacian operator can be simplified to:
\begin{equation}
\nabla^2=\frac{\partial^2}{\partial r^2} +\frac{d_s-1}{r}\frac{\partial}{\partial r},
\label{Laplacen operator in D-dimension media}
\end{equation}
Therefore, we can express the general equation for the heat condition in isotropic fractal media as:
\begin{equation}
\frac{1}{D}\frac{\partial T(r,t)}{\partial t}=\frac{\partial^2}{\partial r^2}{T}(r,t)+\frac{d_s-1}{r}\frac{\partial}{\partial r}T(r,t)+\psi(r,t).
\end{equation}

\section{Results and Simulations}
In what follows, we are going to present analytical formulas and numerical results for two typical transient diffusion cases in fractal dimensions: A) Heat diffusion driven by fixed temperature bias and heat sources; and B) Transient pulsed heat diffusion with random absorbing sinks.

\subsection{Fixed temperature boundaries and heat sources}
Assume the fractal medium boundary is surrounded by heat sinks at the distance $r=R$ with the fixed boundary condition $T(r=R,t)=T_0$ and the initial temperature profile $T(r,t=0)=\mu(r)$, we can obtain the solution:
\begin{eqnarray}
T(r,t)=T_0&+&\sum_{n=1}^{\infty}\frac{2 J_{\frac{d_s}{2}-1}(\frac{\xi_n r}{R})}{r^{\frac{d_s}{2}-1}J_{\frac{d_s}{2}}(\xi_n)^2 }
\int_0^R \big[\frac{\mu(r)-T_0}{R^2}e^{-\frac{\xi^2_n}{R^2}Dt}   \nonumber \\
&+&\frac{\psi(r,t)}{\xi^2_n}(1-e^{-\frac{\xi^2_n}{R^2}Dt})\big]r^{\frac{d_s}{2}}J_{\frac{d_s}{2}-1}(\frac{\xi_n r}{R}) dr,
\label{eq:mainT}
\end{eqnarray}
where $\xi_n$ is the $n$-th zero point of the $d_s/2-1$ fractional order Bessel function $\mathcal J_{d_s/2-1}(\xi_n)=0$. The first part in the integration describes the multi-time-scale relaxation from the initial temperature profile to the fixed boundary temperature due to the heat diffusion. The second part in the integration denotes the temperature raising due to the heat source flux. 

Therefore, we can calculate the heat flux as a function of location by making the gradient $J(r,t)=-\kappa \nabla T(r,t)$:
\begin{eqnarray}
J(r,t)&=& \sum_{n=1}^{\infty}\frac{2\kappa \mathcal J_{\frac{d_s}{2}}(\frac{\xi_n r}{R})\frac{\xi_n}{R}}{\mathcal J_{\frac{d_s}{2}}(\xi_n)^2 r^{\frac{d_s}{2}-1}}
\int_0^R  \big[\frac{\mu(r)-T_0}{R^2}e^{-\frac{\xi^2_n}{R^2}Dt}   \nonumber \\
&+&\frac{\psi(r,t)}{\xi^2_n}(1-e^{-\frac{\xi^2_n}{R^2}Dt})\big]r^{\frac{d_s}{2}}\mathcal J_{\frac{d_s}{2}-1}(\frac{\xi_n r}{R}) dr,
\label{eq:mainJ}
\end{eqnarray}
As the same, the first flux results from the heat diffusion due to the temperature difference between the interior profile and the exterior fixed boundary condition, and the second one results from the inject heat flux due to the source term, respectively. 

Accordingly, using the hyper-sphere volume $V_{d_s}(r)=\frac{\pi^{d_s/2}}{\Gamma(d_s/2+1)}r^{d_s}$ and integrating $E=\int_0^R c\rho T dV_{d_s}=\int_0^R c\rho T(r) \frac{2\pi^{d_s/2}}{\Gamma(d_s/2)}r^{d_s-1} dr$ , we can also calculate the medium's excess energy, $\Delta E=E-c\rho V_{d_s}(R)T_0$, as
\begin{eqnarray}
\Delta E&=&\sum_{n=1}^{\infty}\frac{4c\rho \pi^{\frac{d_s}{2}}R^{\frac{d_s}{2}+1}}{ \Gamma(\frac{d_s}{2})\mathcal J_{\frac{d_s}{2}}(\xi_n)\xi_n}
\int_0^R \big[\frac{\mu(r)-T_0}{R^2}e^{-\frac{\xi^2_n}{R^2}Dt}   \nonumber \\
&+&\frac{\psi(r,t)}{\xi^2_n}(1-e^{-\frac{\xi^2_n}{R^2}Dt})\big]r^{\frac{d_s}{2}}\mathcal J_{\frac{d_s}{2}-1}(\frac{\xi_n r}{R}) dr,
\label{eq:mainE}
\end{eqnarray}

Clearly, the thermal relaxation process is described by the series of time decay factors $\exp(-{\xi_n^2}Dt/{R^2})$. As such, in the long time limit, the characteristic time-scale $\tau_c$ of the heat conduction relaxation is governed by the first term with the smallest exponent $\xi_1^2$, which can be written as 
\begin{equation}
\frac{R^2}{(2+d_s/2)d_s D}<\tau_c=\frac{R^2}{\xi_1^2 D}<\frac{R^2}{2d_s D}.
\label{eq:bound}
\end{equation}
Clearly, increasing $d_s$ will decrease the characteristic time, which means larger dimension can promote the heat diffusion.
Note $\xi_1(d_s)$, as the first zero of the Bessel function $ J_{d_s/2-1}$, is a function of the fractal dimension $d_s$ and has the theorem ~\cite{cite15, cite16}: $\frac{2}{d_s(d_s+4)}<\frac{1}{\xi_1^2}<\sum_{n=1}^{\infty}\frac{1}{\xi_n^2}=\frac{1}{2 d_s}$. 

We can compare it with the Brownian random motion of a single free particle immersed in a thermalized environment in multi-dimension $d_s$. The mean square displacement (MSD) increases linearly with time as $\langle x^2\rangle=2d_sD t$. When replacing the MSD $\langle x^2\rangle$ by the square radius $R^2$, it is clear to see that the upper bound $\tau_c$ in Eq.~(\ref{eq:bound}) is exactly the time for the MSD of the Brownian particle's random walk reaching the square radius. In other words, the diffusion dynamics of heat conduction under fixed temperature boundaries is faster than the random diffusion of Brownian particles. This is understandable, because the former one is essentially a nonequilibrium process driven by external fields, including both the heat sources and temperature boundary conditions with inherent thermal bias, while the latter process is essentially an equilibrium under a uniform thermalized environment with no temperature bias. Therefore, the latter Brownian diffusion is slower with a larger time scale as the upper bound of the time scale of former heat diffusion. 


\begin{figure}
\hspace{-2mm}
\scalebox{0.48}[0.48]{\includegraphics{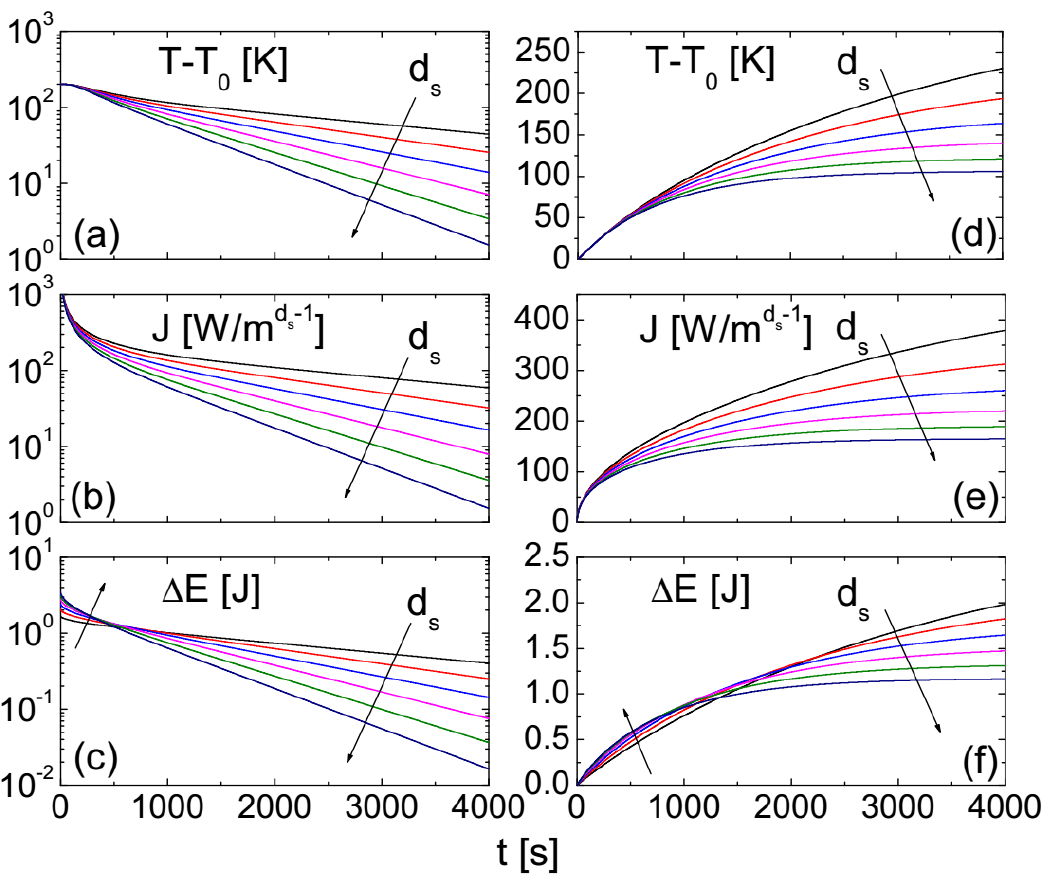}}
\caption{{\bf Transient heat conduction driven by temperature difference (a)(b)(c) or heat source (d)(e)(f).} For conduction driven by temperature difference: (a) Plots of the temperature evolution $T(r,t)-T_0$ at $r=0.6$ m; (b) Plots of the density evolution of the heat flow $q(r,t)$ at $r=1$ m; (c) Plots of the excess energy decay $\Delta E(d_s,t)$. Dimension $d_s$ increases from 1 to 3 with increment of 0.4 in the direction of the arrow. Parameters are $1/D=8.1\times10^3$ s/m$^2$, $R=1$ m, $\kappa=518.52$ W/(m$^{d_s-2} \cdot$K), $c=4.2\times10^3$ J/(kg$\cdot$K), $\rho=10^3$ kg/m$^{d_s}$, $\mu(r)=300$ K, $T_0=100$ K, $\psi(r,t)=0$. For conduction driven by heat source: (d) Plots of the temperature evolution $T(r,t)-T_0$ with $r=0.6$m; (e) Plots of the density evolution of the heat flow $J(r,t)$ with $r=1$m; (f) Plots of the incremental energy $\Delta E(d_s,t)$ Dimension $d_s$ increases from 1 to 3 with increment of 0.4 in the direction of the arrow, where $\mu(r)=T_0=100$ K, $\psi(r, t)=10^3$ K/m$^{d_s-1}$. Other parameters remain the same.
}
\label{fig:Fig2}
\end{figure}


In the following, we will present the associated numerical calculations.
Figure~{\ref{fig:Fig2}}(a)(b)(c) plot the transient heat conduction only driven by temperature difference in the absence of heat source $\psi(r)=0$. The media of larger dimension $d_s$ can promote the heat propagation, so as to dissipate energy faster [see Fig.~{\ref{fig:Fig2}}(a)]. This leads to smaller temperature gradient so as smaller heat flux density on the boundary of the media, as shown in Fig.~{\ref{fig:Fig2}}(b). 
As shown in Fig.~\ref{fig:Fig2}(c), $\Delta E$ also decreases faster as $d_s$ increases. 
This is intuitively because higher dimension means each physical site has more neighbors so that the propagation has more paths to spread out, more efficient and easier.

%

Figure~\ref{fig:Fig2}(d)(e)(f) plot the transient heat conduction only driven by heat sources  $\psi(r,t)\neq 0$. In other words, this case assumes that at the initial moment, the temperature in the media is the same as that in the exterior boundary, which means the uniform heat source exists to persist $\mu(r)=T_0$. 
From Fig.~\ref{fig:Fig2}(d), we see that although with a lower saturated temperature, the temperature saturates faster in media of larger dimensions $d_s$. This indicates that the larger $d_s$ can promote the faster heat conduction. As $d_s$ increases, since the temperature saturates faster to a lower value, the temperature gradient on the boundary is smaller so that the heat flux density is smaller on the boundary of the media [see Fig.~\ref{fig:Fig2}(e)].  
The behaviour of $\Delta E$ is more complicated. From Eq.~(\ref{eq:mainE}), we see that at a short time, the change rate of $\Delta E$ is proportional to $\frac{\pi^{d_s/2}R^{d_s}}{2d_s\Gamma(d_s/2)}$. While in the long time limit, the saturated $\Delta E$ is proportional to $\frac{\pi^{d_s/2}R^{d_s+2}}{(d_s+2)d_s^2\Gamma(d_s/2)}$. (We used $\sum_{n=1}^{\infty}\frac{1}{\xi_n^4}=\frac{1}{2 d_s^2 (d_s+2)}$ in Ref.~\cite{series}.) Correspondingly, as shown in Fig.~{\ref{fig:Fig2}}(f),  the change rate of $\Delta E$ increases as $d_s$ increases, although  the saturated $\Delta E$ decreases.

\subsection{Pulsed heat diffusion with random absorbing sinks}

It is worth noting that many subjects, such as life science and material science, have came across a similar problem: how to get the temperature profile and  energy evolution when an initial heat pulse is excited in a fractal medium. That is to say, at the initial moment, the temperature focused at an spot is much higher than the other part of the media, and may be described by a Dirac delta function. For example, when a cancer tissue has been hit by the focused gamma ray beams, a spot of the cancer tissue can be at a very high temperature. If we can get the temperature profile and energy evolution in the cancer tissue, it will help to adjust the beam intensity or other parameters to achieve the goal of killing the cancer tissue. (Here is a document about using gamma ray knife to treat cancer~\cite{cite17}.) Heat pulse excitation is also applied to materials to detect their intrinsic thermal diffusion and conductivity properties. 

This kind of problem is a specific form of the general problem we have risen above but if we take the Dirac delta function directly into the formula Eq.~(\ref{eq:mainT}), we may not get the final result simply. So we use different scenario and mathematical methods to deal with this problem: apply a heat pulse to the center of a fractal sphere of uniform temperature, which may dissipate if introducing the distributed absorbing heat sinks. 

Assume the heat pulse takes the form of Gaussian distribution ${T_p}\exp(-\pi r^2/a^2)$ with $T_p$ a high temperature. When the length scale $a$ is small enough, the heat pulse may be rewritten as $T_pa(\frac{\exp(-\pi r^2/a^2)}{a})=T_pa\delta(r)$, as a Dirac delta function.
Therefore, denoting $T_0$ the initial ambient temperature and the fixed temperature of heat sinks surrounding the fractal sphere boundary, the initial condition and boundary condition are given by
$T(r,t=0)=T_{0}+T_pa\delta(r)$
and $T(r=R,t)=T_0$.

The temperature profile evolution under a heat pulse can be solved in an analytical form by setting $\psi(r,t)=0$ in Eq.~(\ref{eq:mainT}),  expressed as
\begin{equation}
T(r,t)=T_0+\sum_{n=1}^{\infty}\mathcal{C}_nr^{1-\frac{d_s}{2}}\mathcal J_{\frac{d_s}{2}-1}(\frac{\xi_nr}{R}) e^{-\frac{\xi_n^2}{R^2}Dt}.
\end{equation}
The amplitude $\mathcal{C}_n$ is given by 
\begin{equation}
\mathcal{C}_n=\frac{T_p a^{d_s}}{\pi R^2 \mathcal J_{\frac{d_s}{2}}(\xi_n)^2}\left(\frac{\xi_n}{2\pi R}\right)^{\frac{d_s}{2}-1},
\end{equation}
with an omitted factor $e^{-\frac{\xi_n^2a^2}{4\pi R^2}}\simeq 1$ for small $a\ll R$.
The heat flux profile evolution is obtained accordingly: 
\begin{equation}
J(r,t)=\sum_{n=1}^{\infty}\mathcal{C}_n\kappa\frac{\xi_nr^{1-\frac{d_s}{2}}}{R}\mathcal J_{d_s/2}(\frac{\xi_nr}{R}) e^{-\frac{\xi_n^2}{R^2}Dt}
\end{equation}
By integrating the distribution of temperature profile over the whole volume $V_{d_s}(R)=\frac{2\pi^{d_s/2}}{d_s\Gamma(d_s/2)}R^{d_s}$ of the fractal sphere, we obtain the excess energy dwelling in the fractal medium at time $t$:
\begin{equation}
\Delta E(t) = E-c\rho V_{d_s}T_0= 
\frac{4c\rho a^{d_s}T_p}{2^{\frac{d_s}{2}}\Gamma(\frac{d_s}{2})} \sum_{n=1}^{\infty}\frac{\xi_n^{\frac{d_s}{2}-2}}{\mathcal J_{\frac{d_s}{2}}(\xi_{n})} e^{-\frac{\xi_n^2}{R^2}Dt}.
\label{eq:pulseE}
\end{equation}
As expected, this solution represents the kinetics of the excess energy injected by the heat pulse that decays from the initial one $\Delta E(t=0)=c\rho a^{d_s}T_p$, with a dwelling fraction $f_E(t)={\Delta E(t)}/{\Delta E(0)}$ at time $t$.

Therefore, the characteristic decay time of the heat pulse's energy can be calculated to a simple expression:
\begin{equation}
\tau=\int_{0}^{\infty}dt f_E(t)=\frac{4R^2}{2^{\frac{d_s}{2}}D\Gamma(\frac{d_s}{2})}\sum_{n=1}^{\infty}\frac{\xi_n^{\frac{d_s}{2}-4}}{J_{\frac{d_s}{2}}(\xi_n)}=
\frac{R^2}{2d_sD}.
\end{equation}
Surprisingly, this decay time of the heat pulse's energy conducting in the
fractal medium is just equal to the diffusion time needed for the
Brownian random walk in this medium to reach the distance $R$. 

We note this observation does not conflict with the previous one [see Eq.~(\ref{eq:bound}) and associate discussions] where the nonequilibrium conduction is faster than the equilibrium diffusion. Here, the decay time is for the heat pulse that was excited approximately as a Dirac delta function, which makes the nonequilibrium process well described by the linear response theory, see Ref.~\cite{cite18}.  And as is well known, in the linear response the nonequilibrium properties have connections to the equilibrium properties, such as the equivalence between heat conduction and diffusion~\cite{cite18}, the connection between nonequilibrium transport coefficients and equilibrium flux autocorrelations in Green-Kubo formula~\cite{cite19}. Thus, it is reasonable that the decay time of the heat pulse is equal to the diffusion time of a Brownian motion in the fractal medium.

So far, we have discussed the free heat diffusion in an unperturbed way in a fractal $d_s$-dimensional hyper-sphere of radius $R$ (so as volume $V_{d_s}$) with no additional absorbing boundaries inside the medium. In reality, the medium interior could have thermal radiation spots or heat sinks randomly distributed as absorbing boundaries of thermal energy. Therefore, we assume that the heat sink distribution follows Poisson statistics, so that the probability to obtain no absorbing sinks inside (but on the boundary of) volume $V_{d_s}$ is given by  $p(V_{d_s})=Ce^{-CV_{d_s}}$, where $C$ is the concentration of absorbing heat sinks. By taking into account all possible routes of the heat pulse diffusion, we calculate the mean excess energy of the heat pulse remaining in the medium by averaging the dynamics above [see Eq.~(\ref{eq:pulseE})] over different realizations of $V_{d_s}$, written as:
\begin{equation}
\overline{\Delta E}(t)=\int_{0}^{\infty}p(V_{d_s})\Delta E(t,R(V_{d_s}))dV_{d_s}.
\label{eq:overline deltaE}
\end{equation}
In the long time limit $t\to\infty$, the decay dynamics of heat pulse's energy is governed by the first term of the series (contains $\xi_1$ only), which with the slowest decay dominates the asymptotic behavior. Thus, we can apply the Laplace's method, or say, the saddle-point approximation, to obtain the asymptotic  
\begin{equation}
\overline{\Delta E}(t) \simeq A_{d_s}\left(C^{\frac{2}{d_s}}Dt\right)^{\frac{d_s}{2{d_s}+4}}e^{-B_{d_s}\left(C^{\frac{2}{d_s}}Dt\right)^{\frac{d_s}{{d_s}+2}}},
\label{eq:lifeE}
\end{equation}
where $A_{d_s}$ and $B_{d_s}$ are coefficients given by 
$A_{d_s}=\frac{2^{3-\frac{d_s}{2}}c\rho a^{d_s}T_p\left[{\pi^{{d_s}+1}\xi_1^{{d_s}^2/2-4}}/{\Gamma({d_s}/2)^{{d_s}+3}}\right]^{\frac{1}{{d_s}+2}}}{\sqrt{2+d_s}\mathcal{J}_{d_s/2}(\xi_1)}$,
$B_{d_s}=\frac{2+{d_s}}{{d_s}}\left[\frac{(\xi_1\sqrt{\pi})^{d_s}}{\Gamma({d_s}/2)}\right]^{\frac{2}{{d_s}+2}}$.  
This asymptotic form clearly shows a non-exponential decay behavior.
More importantly,  the time-dependent decay behavior depends mainly on two undetermined parameters: the dimension of the fractal medium ${d_s}$ and the concentration of  absorbing heat sinks  $C$, with thermal diffusivity $D$ just rescaling the time, which all can be fitted out from the time-dependent experimental measures. 

Similarly, taking into account all possible routes of the heat pulse diffusion, we calculate the mean decay time by averaging over the probability distribution of $V_{d_s}$, as:
\begin{equation}
\overline\tau=\int_{0}^{\infty}dV_{d_s}p(V_{d_s})\int_{0}^{\infty}dtf_E(t)=\frac{\Gamma(\frac{2}{d_s})\Gamma(1+\frac{d_s}{2})^{\frac{2}{d_s}}}{\pi d_s^2 DC^{\frac{2}{d_s}}}.
\label{eq:lifeT}
\end{equation}
The same, by measuring the average decay time for different heat sink concentration $C$, we can also fit out the fractal dimension $d_s$ and the thermal diffusivity $D$. It is worth noting that $\overline\tau$ has a nontrivial dependence on dimension for difference $C$. For example, only at small $C$, increasing $d_s$ will decrease the decay time, which means larger dimension can promote the heat diffusion; while at large $C$, things are reversed. 

\begin{figure}
\vspace{2mm}
\scalebox{0.4}[0.4]{\includegraphics{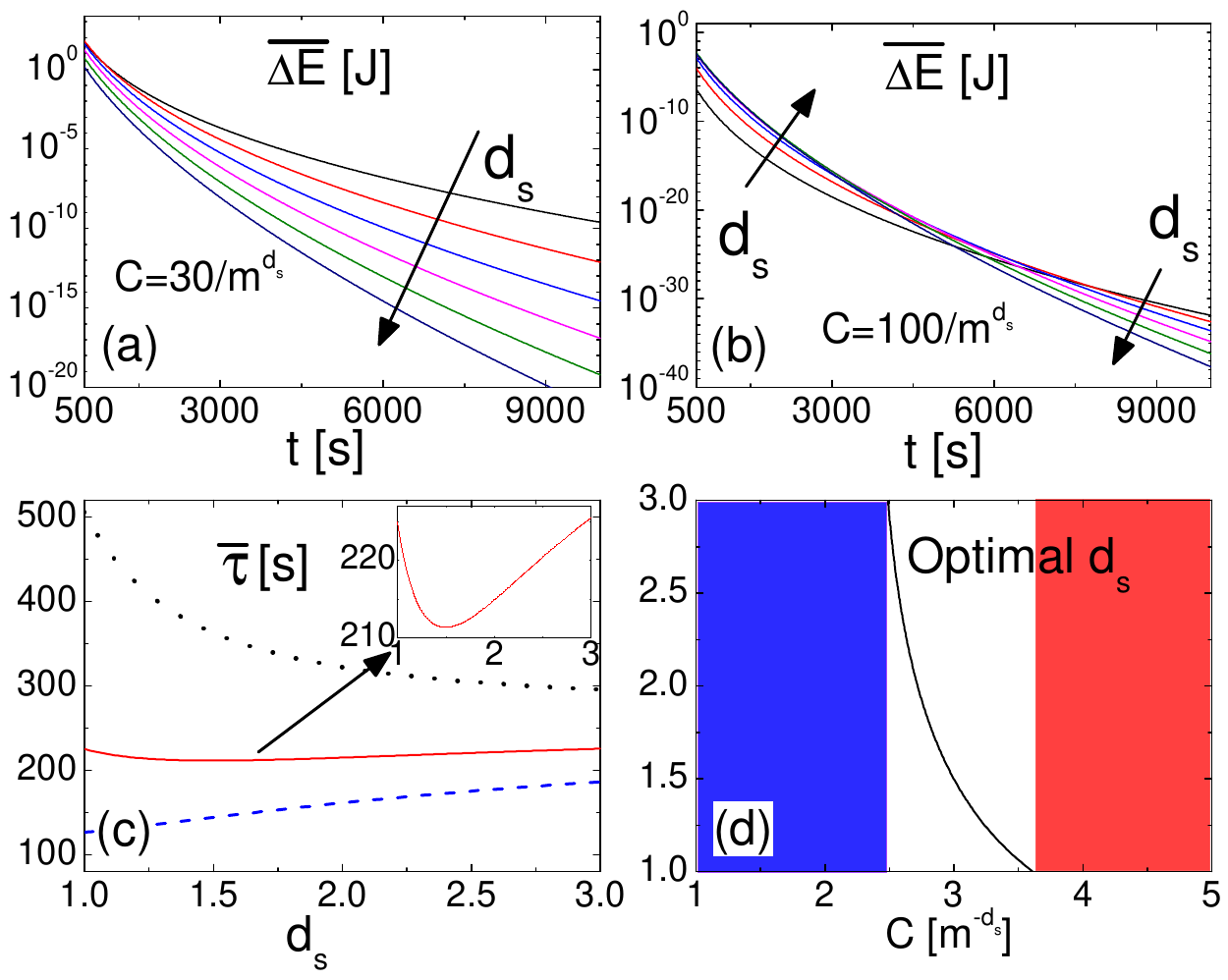}}
\vspace{-2mm}
\caption{
{\bf Heat pulse diffusion with random absorbing sinks in fractal media.} (a)(b) Plots of $\overline{\Delta E}(t)$ from Eq.~(\ref{eq:lifeE})  for $C=30/$m$^{d_s}$ and $C=100/$m$^{d_s}$, respectively, show complicated dimension dependences. Dimension $d_s$ increases from 1 to 3 with increment of 0.4 indicated by the arrow direction; (c) Plots of the mean decay time $\overline\tau(d_s)$ from Eq.~(\ref{eq:lifeT}) with $C=2/$m$^{d_s}$ in upper dot black line, $3$/m$^{d_s}$ in middle solid red line, $4$/m$^{d_s}$ in lower dash blue blue; (d) Optimal dimension $d_s$ with minimal decay time for efficient heat absorption. $T_0=100$ K, $\psi(r,t)=0$, $T_p=10^3$ K, $a=0.01$ m. Other parameters are the same as before. Note, we limit $d_s$ between 1 and 3, considering reality.}
\label{fig:Fig3}
\end{figure}

Also, we  present here the associated numerical calculations.  
For the case of pulsed heat diffusion, the temperature evolution $T(r,t)-T_0$ and the density evolution of the heat flow $J(r,t)$ show similar behaviors as those in Fig.~{\ref{fig:Fig2}}(a)(b). The excess energy decay $\Delta E(d_s,t)$ is faster as $d_s$ increases.  Therefore, we do not show them here repeatedly. 
For the heat pulse with random absorbing sinks, we plot the numerical results of $\overline{\Delta E}(t)$ and $\overline{\tau}(d_s)$ in Fig.~\ref{fig:Fig3}, by applying sharp Gaussian pulse with $T_p=10^3$ K and $a=0.01$ m.
Figure~\ref{fig:Fig3}(a)(b) show clearly the non-exponential behaviors of the mean excess energy decay $\overline{\Delta E}(t)$ for heat sink concentration $C=30/$m$^{d_s}$ and $C=100/$m$^{d_s}$, respectively. They have very complicated dimension dependences, as we can see from Eq.~(\ref{eq:lifeE}) . 

The mean decay time $\overline{\tau}$, [see Eq.~(\ref{eq:lifeT})], also has a nontrivial dependence on $d_s$, which is affected by the concentration $C$ of absorbing heat sinks. When $C$ is relatively large, $\overline{\tau}$ increases as $d_s$ increases, as the dash line in Fig.~\ref{fig:Fig3}(c). When $C$ is relatively small, $\overline{\tau}$ decreases as $d_s$ increases, as the dot line in Fig.~\ref{fig:Fig3}(c). If $C$ is set to be an intermediate value in the middle, then $\overline{\tau}$ is allowed to obtained a minimal value with an optimal dimension $d_s$, where the heat absorption is more efficient, see the solid line in Fig.~\ref{fig:Fig3}(c) with enlarged view in the inset. The optimal  $d_s$ is depicted in Fig.~\ref{fig:Fig3}(d), which indicates that at intermediate $C$, the optimal dimension for efficient heat absorption is closed to the solid line. At lower sink concentrations, the larger dimension ($d_s\rightarrow 3$) the better, while at higher sink concentrations, the lower dimension ($d_s \rightarrow 1$) the better. 

\section{Discussions}

As we have noted in the beginning,  the fractal structures studied at present are beyond the mesoscopic level and the unit size of the fractal structure is above the micrometer instead of at nanoscale, so that Fourier's law is valid. To include the non-Fourier cases, one can generalize the master equation~\ref{common equation} (without loss of generality, we omit the source term): $\frac{\partial T}{\partial t}=D\nabla^2 T(\vec r, t)$ with constant heat diffusion coefficient $D={\kappa/(c\rho)}$ to a non-Markov diffusion equation with memory:
\begin{equation}
\frac{\partial T(\vec r, t)}{\partial t}=\int^{t}_0K(t-t')\nabla^2 T(\vec r, t')dt'.
\end{equation}
The retarded function $K(t)$ is called the memory kernel, which plays the important role in diffusion dynamics so that the future will not only depend on the present state but also on the history. Some special kernels will lead to familiar equations as follows:
\begin{eqnarray}
\text{No memory}: K(t)&=&2D\delta(t), \nonumber \\ \text{Diffusion equation}: \frac{\partial T(\vec r, t)}{\partial t}&=&D\nabla^2 T(\vec r, t);   \\
\text{Full memory}: K(t)&=&v_s^2\Theta(t), \nonumber \\ \text{Ballistic wave equation}: \frac{\partial^2 T(\vec r, t)}{\partial t^2}&=&v_s^2\nabla^2 T(\vec r, t);   \\
\text{Decaying memory}: K(t)&=&v_s^2e^{-v_s^2 t/D}, \nonumber \\ \frac{1}{v_s^2} \frac{\partial^2 T(\vec r, t)}{\partial t^2}+\frac{1}{D} \frac{\partial T(\vec r, t)}{\partial t}&=&\nabla^2 T(\vec r, t).
\end{eqnarray}
A completely no memory kernel $K(t)\sim\delta(t)$ leads to the normal diffusion process, while a full memory of the history $K(t)\sim\Theta(t)$ gives the ballistic wave equation. In between, the decaying memory kernel leads to the telegraph equation that combines the diffusion and ballistic wave equation. By applying the fractional dimension Laplacian operator $\nabla^2=\frac{\partial^2}{\partial r^2} +\frac{d_s-1}{r}\frac{\partial}{\partial r}$, we can generalize the present discussions to the general diffusion process with non-Markovian memory kernels. 

So far, we restricted ourself to the constant diffusion coefficient $D={\kappa/(c\rho)}$. We tried to fix the diffusion constant to merely see the pure dimension effect from varying $d_s$. This kind of scheme also follows the treatment in Refs.~\cite{cite10,cite12}, where they also consider the mass density $\rho$, thermal conductivity $\kappa$, and so as diffusion coefficient $D$ as a constant, when dealing with the problem about heat conduction in fractal media. Nevertheless, we need to note that in general the properties $\kappa, c, \rho$ will have rich dependences on the dimension $d_s$. Once the explicit dimension dependence is known, we can replace those constants as  functions  of $d_s$, to see more rich and complicated heat conduction behavior in fractional dimension. 

Moreover, given spatial dependent system parameters, the fractional dimensional diffusion equation with different memory kernels, can be straightforwardly applied to the transformation thermodynamics~\cite{jphuang}, to design the transformed thermal cloaking, camouflage and so on,  with fractional dimensions and anomalous non-Fourier thermal behaviors.

\section{Conclusion}
In this paper, we have described the transient heat conduction in fractal media in the framework of calculus in fractional dimension space. We have studied the influence of dimension to the evolution of the temperature distribution, the density of the heat flux and the excess energy, and several examples are analyzed to illustrate the results. We have found that in general larger dimension can promote heat propagation, but may with a complicated dependence on the system parameters. A special case for the heat pulse has been considered. With randomly distributed heat sinks in the media, we have obtained a non-exponential decay behavior of the excess energy, and the time-dependent decay behavior depends mainly on two undetermined parameters: the dimension of the fractal medium ${d_s}$, and the concentration of absorbing heat sinks $C$. At lower sink concentrations, the large dimension promotes the heat absorption, while at higher sink concentrations, the lower dimension the better.  An optimal dimension for efficient heat absorption emerges for intermediate $C$. By experimentally measuring the time-dependent kinetics, one may fit out the dimension of the fractal medium ${d_s}$, the concentration of  absorbing heat sinks  $C$, and the thermal diffusivity $D$.

Our results may have implications in material science, life science and medical science to describe transient heat conduction in fractal media like porous media, living tissue and composite. We hope they can be used to guide the design of thermal device, controlling transient heat conduction in ubiquitous fractal media, such as porous, composite, networked materials.

\section*{appendix}
\subsection{The simple derivation of the general solution}
The heat diffusion in the fractal medium is expressed as:
\begin{equation}
\frac{1}{D}\frac{\partial T(r,t)}{\partial t}=\frac{\partial^2}{\partial r^2}{T}(r,t)+\frac{d_s-1}{r}\frac{\partial}{\partial r}T(r,t)+\psi(r,t).
\end{equation}
Suppose that this problem's boundary conditions and initial conditions are $T(r,t=0)=\mu(r), T(r=R,t)=T_0$.
Separate this temperature in its mathematical form $T=T_1+T_2+T_0$, for the $T_1$ and $T_2$, the following equations are satisfied respectively:
\begin{eqnarray}
\left(\frac{\partial^2}{\partial r^2}+\frac{d_s-1}{r}\frac{\partial}{\partial r}\right){T_1}-\frac{1}{D} \frac{\partial T_1}{\partial t}&=&0,  \\
\left(\frac{\partial^2}{\partial r^2}+\frac{d_s-1}{r}\frac{\partial}{\partial r}\right){T_2}-\frac{1}{D} \frac{\partial T_2}{\partial t}&=&-\psi(r,t).
\end{eqnarray}
The $T_1,T_2$ satisfy the following initial conditions and  boundary conditions respectively: $T_1(r,t=0)=\mu(r)-T_0, T_2(r,t=0)=0; T_1(r=R,t)=0, T_2(r=R,t)=0$.

We solve the first equation by using the variable separation $T_2(r,t)=R_2(r)f_2(t)$, then we get these two eigenvalue equations:
\begin{eqnarray}
&&\frac{d^2 R_2}{dr^2}+\frac{d_s-1}{r}\frac{dR_2}{dr}+\lambda^2 R_2=0,     \\  
&&\frac{df_2}{dt}=-{\lambda^2}D f_2.
\end{eqnarray}
Because of the boundary condition, we can confirm that the eigenvalue-related parameter $\lambda$ we introduce above, is
\begin{equation}
\lambda_n=\frac{\xi_n}{R},
\end{equation}
where $\xi_n$ is the nth zero point of the $(d_s/2-1)$ order Bessel function $\mathcal{J}_{d_s/2-1}=0$. 
Solve all the equations, then we get the general solution of the first equation:
\begin{eqnarray}
T_1(r,t)=&&\sum_{n=1}^\infty[A_n J_{|1-\frac{d_s}{2}|}(\lambda_n r)+B_n J_{-|1-\frac{d_s}{2}|}(\lambda_n r)]  \nonumber \\
&&\times r^{1-\frac{d_s}{2}} e^{-\lambda^2_n Dt},
\end{eqnarray}

For the second equation, we use impulse theorem to solve it. As people often do in solving this kind of typical mathematical physics problems, we suppose that:
\begin{equation}
T_2(r,t)=\int_{0}^{t}v_1(r,t;\tau)d\tau.
\end{equation}
What the equation $v_1$ satisfies is very similar to what the equation $T_1$ satisfies, so we can easily write down its general solution. Then we make an integral, and can write down the the general solution of $T_1$:
\begin{eqnarray}
T_2(r,t)=&&\sum_{n=1}^\infty[C_n J_{|1-\frac{d_s}{2}|}(\lambda_n r)+
D_n J_{-|1-\frac{d_s}{2}|}(\lambda_n r)]   \nonumber \\
&&\times r^{1-\frac{d_s}{2}}\frac{1}{D \lambda^2_n} (1-e^{-\lambda^2_n Dt}).
\end{eqnarray}

Considering the initial conditions, we can finally write down the special solution of $T=T_0+T_1+T_2$, with
\begin{eqnarray*}
T_1&=&\sum_{n=1}^{\infty}\frac{2e^{-\frac{\xi^2_n D}{R^2}t}J_{\frac{d_s}{2}-1}(\frac{\xi_n r}{R})}{r^{\frac{d_s}{2}-1}J_{\frac{d_s}{2}}(\xi_n)^2 }   \\
&\times&\int_0^R \frac{\mu(r)-T_0}{R^2} r^{\frac{d_s}{2}}J_{\frac{d_s}{2}-1}(\frac{\xi_n r}{R}) dr     
\nonumber \\
T_2&=&\sum_{n=1}^{\infty}\frac{2(1-e^{-\frac{\xi^2_n D}{R^2}t})J_{\frac{d_s}{2}-1}(\frac{\xi_n r}{R})}{r^{\frac{d_s}{2}-1}J_{\frac{d_s}{2}}(\xi_n)^2 \xi^2_n}     \\
&\times&\int_0^R \psi(r,t)r^{\frac{d_s}{2}}J_{\frac{d_s}{2}-1}(\frac{\xi_n r}{R}) dr 
\end{eqnarray*}
which is exactly the Eq.~(\ref{eq:mainT}). 
Then we can calculate the heat flux by means of making a differential [see Eq.~(\ref{eq:mainJ})], and calculate the energy by making an integral [see Eq.~(\ref{eq:mainE})].

\subsection{The saddle-point method to get the asymptotic result}

In order to get the result of the $\overline{\Delta E}(t)$ [see Eq.~(\ref{eq:overline deltaE})], we use a mathematical method called saddle-point method. Because the dominate factor that influences the asymptotic  result of the integral is the first term of the series, so we can calculate only the first term, and use the result to approximate the accurate result. We can provide the concrete form of the integral formula:
\begin{eqnarray}
\overline{\Delta E}(t)&&=\frac{8C\pi^{\frac{d_s}{2}}(\xi_1)^{\frac{d_s}{2}-2}}{2^{\frac{d_s}{2}}[\Gamma(\frac{d_s}{2})]^2\mathcal{J}_{\frac{d_s}{2}}(\xi_1)}   \nonumber \\
&&\times\int_0^{\infty} e^{-(\frac{\xi_1}{R})^2 Dt-\frac{C\pi^{\frac{d_s}{2}}R^{d_s}}{\Gamma(1+\frac{d_s}{2})}}R^{d_s-1}dR
\end{eqnarray}
Then we suppose that $\alpha=(\xi_1)^2 Dt$, $\beta=\frac{C\pi{\frac{d_s}{2}}}{\Gamma(1+\frac{d_s}{2})}$, then the integral (excluding the coefficient) can be written as $\int_0^{\infty}R^{d_s-1}e^{-\frac{\alpha}{R^2}-\beta R^{d_s}}dR=\int_0^{\infty}g(R)\exp[zh(R)]dR$. In this formula, $g(R)=R^{d_s-1}$, $h(R)=-\frac{\alpha}{R^2}-\beta R^d_s$.

Suppose that 
\begin{equation}
f(z)=\int_0^{\infty} g(R)e^{zh(R)}dR.
\end{equation}
By using steepest descent method, we can know that if $R$ is always a real number, the $f(z)$ can be approximated by the beneath expression~\cite{cite20}:
\begin{equation}
f(z)\sim i\sqrt{\frac{2\pi}{zh''(R_0)}}g(R_0)\exp[zh(R_0)].
\end{equation}
In the present circumstances, $z=1$, then the integral's approximate expression is $i\sqrt{\frac{2\pi}{h''(R_0)}}g(R_0)\exp[h(R_0)]$. The $R_0$ in the formulas is the zero point of the function $\partial h(R)/\partial R=0$. In this problem it is $R_0=(\frac{2\alpha}{\beta d_s})^{\frac{1}{2+d_s}}$. We can substitute this result in the approximate expression of the integral, then we can get:
\begin{equation}
f(1)\sim \sqrt{\frac{\pi}{\alpha(2+d_s)}}(\frac{2\alpha}{\beta d_s})^{\frac{d_s+1}{d_s+2}}\exp[-\beta \frac{2+d_s}{2}(\frac{2\alpha}{\beta d_s})^{\frac{d_s}{d_s+2}}].
\end{equation}
Then we substitute the coefficient and the concrete expression of $\alpha$ and $\beta$ in the above formula, and we get the result Eq.~(\ref{eq:lifeE}) in the main text.

\acknowledgements
Our work is supported by the National Natural Science Foundation of China with grant No. 11775159, the Fundamental Research Funds for the Central Universities, and the Opening Project of Shanghai Key Laboratory of Special Artificial Microstructure Materials and Technology.

\end{document}